\documentclass[9pt,twocolumn,twoside]{osajnl}
\journal{ol} % Choose journal (ao, aop, josaa, josab, ol, optica, pr)

\setboolean{shortarticle}{true}
\title{Spectral caustics of high-order harmonics in one-dimensional periodic crystals}

\author[1]{Jiaxiang Chen}
\author[2,3]{Qinzhi Xia}
\author[1,4]{Libin Fu}

\affil[1]{Graduate School of China Academy of Engineering Physics, No. 10 Xibeiwang East Road, Haidian District, Beijing, 100193, China}
\affil[2]{Institute of Applied Physics and Computational Mathematics, Beijing 100088, China}
\affil[3]{e-mail: xia\_qinzhi@iapcm.ac.cn}

\affil[4]{e-mail: lbfu@gscaep.ac.cn}

\begin{abstract}
	We theoretically investigate the spectral caustics of high-order harmonics in solids. We analyze the one-dimensional model of the high-order harmonic generation (HHG) in solids and find that apart from the caustics originating from the van Hove singularities in the energy band structure, another kind of catastrophe enhancement also emerges in solids when the different branches of electron-hole trajectories generating high-order harmonics coalesce into a single branch. We solve the time-dependent Schr\"odinger equation in terms of the periodic potential and demonstrate the control of this kind of singularity in HHG with the aid of two-color laser fields. The diffraction patterns of the harmonic spectrum near the caustics agree well with the interband electron-hole recombination trajectories predicted by the semiconductor semiclassical equation. This work is expected to improve our understanding of the HHG dynamics in solids and enable us to manipulate the harmonic spectrum by adjusting the driving field parameters.
\end{abstract}

\setboolean{displaycopyright}{true}

\begin{document}
	
	\maketitle
	
	In the last decade, growing evidence has emerged that high-order harmonic generation (HHG) in solids has potential applications in compact extreme ultraviolet attosecond light sources \cite{GhimireNP2010,VampaNP2017,GargNP2018, KruchininRMP2018, YangNP2019} and the optical detection of the electronic properties of materials \cite{LuuN2015,SivisSC2017,YouNC2017}, such as the energy band structure \cite{VampaPRL2015,LaninO2017,LiPRL2019} and berry curvature \cite{LuuNc2018,LiuNP2017}. Therefore, much interest exists regarding the study of the characteristics of HHG in solids, such as the plateau structure \cite{NdabashimiyeN2016,DuPRA2016,YuPRA2020}, crystal orientation dependence \cite{YouNP2017,WuPRA2017,JiangPRL2018,ZhangPRB2019,YouOL2019}, and spectral interference \cite{KimACS2019,HohenleutnerN2015,DuPRA2018}. In this work, we focus on the spectrum caustic structure originating from some singularity during the HHG processes in solids. This study on the spectrum caustics is expected to benefit enhancement of the intensity of harmonic signals and  manipulation of the harmonic spectrum. In addition, it will deepen our understanding of the dynamics of electrons and holes in solids.
	
	Regarding HHG in gases, the singularity, where the long and short trajectories of a recollision electron coalesce into a single trajectory, has long been known to create some caustic enhancement in the spectrum.
	Recently, the van Hove singularities \cite{UzanNP2020} of the energy band structure in solids have been reported to result in the spectral caustics in HHG. The singular structure in the spectrum enhances the HHG intensity and reveals information regarding the atoms or solids, such as electronic wavefunctions \cite{AzouryNP2019} and the strong-field modification of the band structure \cite{UzanNP2020}. On the other hand, one of the main mechanisms of HHG in solids is the recombination of the tunneled electron and the hole, which is similar to the three-step model widely accepted in gas HHG processes \cite{CorkumPRL1993,LewensteinPRA1994,VampaPRB2015}. Therefore, whether some kinds of singularities in gas HHG processes remain in the solid-related processes is of great interest. Specifically, we desire to search for spectral caustic structures in solids similar to those in gases \cite{RazNP2011,FaccialaPRL2016,HamiltonPRA2017,PisantyJPP2020}.
	
	The spectrum caustics can be analyzed within the framework of catastrophe theory \cite{ConnorMP1976,BerryPO1980}. During the theoretical treatment of the HHG process, we usually need to evaluate an integral (e.g., \eqref{Jer} in the following) with an exponential function in the integrand. The exponent varies rapidly enough that we can use the stationary phase method to treat the integral under the assumption that all the saddle points are isolated \cite{XiaPRL2018}. However, the high-order derivation of the rapidly varied exponent can be zero, making the assumption invalid; hence, we encounter the singularity, which can be categorized and treated by catastrophe theory. Previous works on HHG in solids \cite{UzanNP2020} have noted that the gradient of the energy band emerges in the second-order derivative of the exponent and concluded that some spectrum caustics originate from the van Hove singularities. Our present work simplifies the formula of the derivative of the exponent and identifies another kind of caustic structure.
	
	This article is based on the two-band model of HHG in solids \cite{VampaPRL2014,VampaPRB2015,LiuPRA2016}. To begin, we propose that the main contribution to the harmonic signal comes from the interband transition when the harmonic energy is above the band gap \cite{VampaPRL2014}. The harmonics can be expressed as:
	\begin{equation}\label{Jer}
		J_{er}(\omega)=\int_{BZ}dk\int_{-\infty}^{\infty} dt \int_{-\infty}^{t}dt' g(k,t',t) e^{-iS(k,t',t)+i\omega t}+c.c.,
	\end{equation}
	where BZ represents the first Brillouin zone and the semiclassical action $S$ can be expressed as:
	\begin{equation}
		S(k,t',t)=\int_{t'}^{t} \varepsilon_{g}(k+A(\tau)-A(t)) d\tau,
	\end{equation}
	with $A(t)$ being the vector potential and $\varepsilon_{g}(k)$ being the band gap at the crystal momentum $k$. Here, we neglect the dephasing effect, and atomic units are used unless otherwise specified. In contrast to the quickly oscillating exponent, $g(k,t',t)$ represents a slowly varying term. Hence, we can use the stationary phase approximation to solve \eqref{Jer}. Then, we have:
	\begin{equation}\label{JerSP}
		J_{er}(\omega)\approx \sum_{k_{st}} (2\pi)^{3/2}e^{i\theta}\frac{g(k_{st},t'_{st},t_{st})e^{-iS(k_{st},t'_{st},t_{st})+i\omega t_{st}}}{\sqrt{|S^{''}(k_{st},t'_{st},t_{st})|}}+c.c.,
	\end{equation}
	where $S^{''}(k_{st},t'_{st},t_{st})$ is the Hessian matrix of the classical action $S(k_{st},t'_{st},t_{st})$ at the saddle point. $k_{st}$ is the crystal momentum with which the electron and the hole recombine, while $t'_{st}$ and $t_{st}$ represent the birth and recombination time of electron-hole pairs, respectively. The phase $\theta$ relies on the angle from which the integral contour passes through the saddle point. According to the conditions on the saddle point $(k_{st},t'_{st},t_{st})$,$k_{st}$, $t_{st}$ and the harmonic energy $\omega$ can be estimated as functions of the birth time $t'_{st}$.	
	
	For simplicity, a one-dimensional two-band model is used here. After some reductions, we find that the determinant of the Hessian matrix of the classical action $S$ can be simplified as:
	\begin{equation}\label{S''}
		|S^{''}|=-\nabla_k \varepsilon_{g}(\kappa_{t'_{st}})\nabla_k \varepsilon_{g}(k_{st})  \frac{d\omega}{dt'_{st}},
	\end{equation}
	where $\kappa_{\tau}=k_{st}+A(\tau)-A(t_{st})$. Details see in Sec. 1 of Supplement 1. 
	
	Interestingly, in contrast to the single kind of HHG spectrum caustics previously identified in solids, we find three factors in \eqref{S''}. The first two terms are the relative velocities of electron-hole pairs at birth time $t'_{st}$ and recombination time $t_{st}$, respectively. Specifically, we notice that when the second term, i.e., $\nabla_k \varepsilon_{g}(k_{st})$, reaches $0$, strong enhancement emerges in Eq. (\ref{JerSP}), as discussed in \cite{UzanNP2020}, which corresponds to the van Hove singularity in the energy band structure of solids. Instead, in this paper, we focus on the third factor, $d\omega/dt'$, which is expressed as:
	\begin{equation}\label{wt}
		\frac{d\omega}{dt'_{st}}=[a(k_{st})+bF(t_{st})]F(t'_{st})-a(\kappa_{t'_{st}})F(t_{st}),
	\end{equation}
	where $a(\kappa_{t'_{st}})=\nabla_k \varepsilon_{g}(\kappa_{t'_{st}})$, $b=\int_{t'_{st}}^{t_{st}}\nabla^2_k \varepsilon_{g}(\kappa_{\tau})d\tau$ and $F(t)=-\partial A(t)/\partial t$.
	$d\omega/dt'$ reaches $0$ at the extreme of the function $\omega(t'_{st})$. For solids in a monochromatic laser field, the extreme can be generated when the long and short electron-hole trajectories merge together, similar to that in gases \cite{RazNP2011,VampaPRB2015}.
	
	Compared to the first two factors[see \eqref{S''}], which rely on the local structure of the energy band, the third item discussed here is jointly determined by the driving field and the band structure.
	This procedure allows us to control the position and intensity of the HHG spectrum caustic with the aid of the external fields. In the following, this is illustrated using two-color laser fields. 
	The method we used to explore the HHG caustics is similar to Ref. \cite{RazNP2011}. Different to the gas-phase HHG in the reference, here we focus HHG process in periodic crystals.

	As a simple illustration, we applied a two-color fields to a one-dimensional periodic system \cite{KruchininRMP2018}. In our numerical simulations, we considered the multielectron dynamics in the one-dimensional model under the independent electron approximation. We solved the TDSE for each electron under the velocity gauge and dipole approximation \cite{IkemachiPRA2017}:
	\begin{equation}\label{TDSEV}
		i\frac{\partial}{\partial t}\psi_{nk}(x,t)=\left[ \frac{1}{2}\left(p+A(t)\right)^2+V(x) \right]\psi_{nk}(x,t),
	\end{equation}
	where $V(x)=-V_0[1+\cos(2\pi x/a_0)]$ is a Mathieu-type potential with $a_0=8\ a.u.$ and $V_0=0.37\ a.u.$. The minimum band gap between the valence and conduction band at this potential is $0.15\ a.u.(4.2\ eV)$. $A(t)$ is the vector potential, with $A(t)=\frac{F_0}{\omega_0}(\sin (\omega_0 t)+\frac{R}{2}\sin(2\omega_0 t+\varphi))f(t)$, where $f(t)$ is the $\sin^4$ envelope and $R$ represents the strength of the second harmonic field compared to the fundamental one. Without loss of generality, the selected parameters of the fundamental field are $F_0=0.0034$ a.u. (corresponding to the peak intensity of $0.4\ \mathrm{TW/cm^2}$) and wavelength $\lambda=3200$ nm to confine the electron motion on the conduction band to the first Brillouin zone. In accordance with \eqref{Jer}, we sum up all the induced currents contributed by the electrons initially fully occupying the highest valence band. 
	
	\begin{figure}[tbp]
		\centering
		\includegraphics[width=0.95\columnwidth]{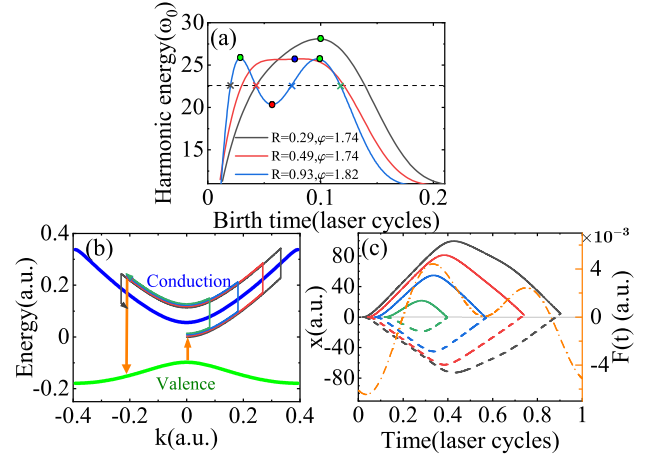}
		\caption{
			(a) High harmonic photon energy as a function of the birth time (ionization time) of the electron-hole pair for three cases: only one maximum (black line), two maxima and one minimum (blue line), and three extreme points merged into one  (red line). The green and red circles denote the maximum and minimum, respectively, while the blue circle is the mergence of two maxima and one minimum. The four crosses represent the birth times of the electron-hole pairs generating photons with the same energy and their trajectories in (b) momentum and (c) real spaces are described in the same color lines.
			(b)  Momentum-space electron trajectories in the band diagram. (c) Real-space trajectories of electrons(solid line) and holes(dash line). The electric field is plotted in orange dash-dotted line.
		}
		\label{figure01}
	\end{figure}

	Based upon the semiclassical motion of electrons and holes \cite{VampaPRB2015,IkemachiPRA2017} in the two-color laser fields, generally four electron-hole pairs with the same recombination energy can be generated in half a laser cycle. These pairs finally recombine and release photons with the same frequency. In Fig. \ref{figure01}(a), we plot the functions of the harmonic energy versus the birth time of electron-hole pairs for different R values. When the second harmonic field is strong, e.g., $R\approx 0.93$, three extremes, i.e., two maximums (green circles) and one minimum (red circle), emerge on the blue line. The four crosses represent the birth times of the electron-hole pairs generating photons of the same energy.
	
	The trajectories of these electron-hole pairs in momentum space and real space are plotted in Fig. \ref{figure01}(b) and \ref{figure01}(c), respectively. In Fig.\ref{figure01}(b), the electrons in the valence band are excited by the laser field and driven forward and backward at the conduction band, while the holes are left in the valence band. When the trajectories of the electrons (solid lines) and holes (dashed lines) recollide in real space, as in Fig.\ref{figure01}(c), the pair recombines and emits the high harmonic photon. Compared to the three-step model for atoms or molecules, in Fig. \ref{figure01}(c), we notice the departure of the recollision position from the origin as a result of the nonquadratic band structure in solids.

	The singularities in Fig. \ref{figure01}(a) imply $d\omega/dt'_{st}=0$ in Eq. (4), which strongly enhances the yield of the harmonics. The singularities can be modulated by the external fields. When the intensity of the second harmonic field weakens, the three extremes on the blue line in Fig. \ref{figure01}(a) get close and merge into a high-order singularity, such as the blue circle on the red line when $R\approx 0.49$. Finally, when $R$ decreases, the laser field is completely dominated by the fundamental field. Only two branches of trajectories remain, i.e., long and short trajectories, which are separated by a single maximum, as demonstrated by the black lines in Fig. \ref{figure01}(a).

	\begin{figure}[ht!]
		\centering
		\includegraphics[width=0.9\columnwidth]{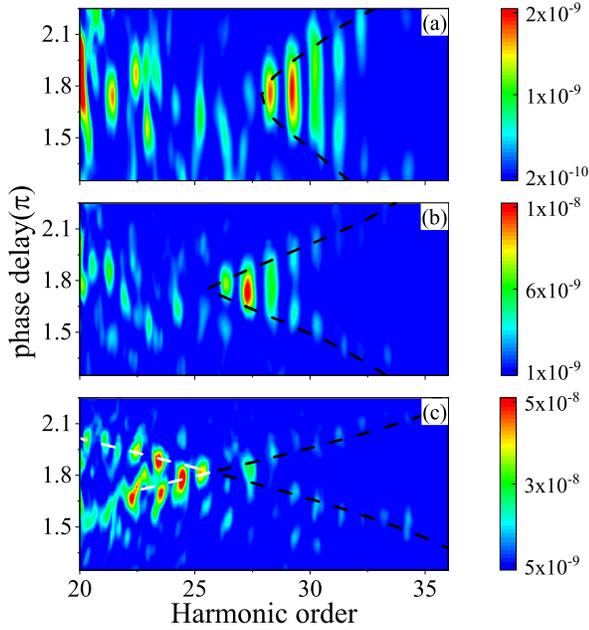}
		\caption{HHG spectra as a function of the phase delay $\varphi$ for $R\approx 0.29$ (a), $R\approx 0.49$ (b) and $R\approx 0.93$ (c), respectively. The dashed lines represent the position of spectral caustics predicted by semiclassical calculations.}
		\label{figure02}
	\end{figure}
	
	By analyzing \eqref{wt}, we found that three main control parameters exist: harmonic energy $(\omega)$, ratio $(R)$ and phase delay $(\varphi)$. Therefore, according to the classification of catastrophe theory \cite{ConnorMP1976,BerryPO1980}, the caustic structure is swallowtail. Figures \ref{figure02}(a)-\ref{figure02}(c) show the swallowtail diffraction pattern for different $R$ and the dashed lines are the caustics zone predicted by \eqref{wt}.
	
	Firstly, when $R$ is weak, we can see a singular enhancement pattern in the HHG spectrum in Fig. \ref{figure02}(a). The enhancement corresponds to the maximum of the returning energy for each phase delay $\varphi$, as demonstrated by the green circle on the black line in Fig. \ref{figure01}(a). When $R$ increases as in Fig. \ref{figure02}(b),  a brighter point emerges at the 27th harmonic and $\varphi\approx 1.74\ \pi$. The point represents the vicinity of the swallowtail point (blue circle) on the red line of Fig. \ref{figure01}(a). Finally, when the intensity of the second harmonic field is strong enough to compare with the fundamental field, a new singular enhancement pattern appears in the direction opposite to those in Figs. \ref{figure02}(a) and \ref{figure02}(b). The new pattern demonstrates the second maximum on the blue line in Fig. \ref{figure01}(a). The diffraction patterns predicted by the semiconductor semiclassical equation are illustrated by the dashed lines in Fig. \ref{figure02}. The simulation results agree well with theoretical predictions, except that the theoretical harmonic orders of the spectral caustics are lower than those in simulations. This small deviation may be partly attributed to the neglect of imaginary part of birth time $t'_{st}$ during the semiclassical trajectories calculation \cite{VampaPRB2015}. 
	
	Additionally, some secondary enhancement structures with lower energy have been displayed in Fig.\ 2(a) and (b). These structures have not been predicted by Eq. (4) and (5). From some extensive calculations, we find that they can be influenced by the transition dipole moment, dephasing time, and laser pulse duration (see details in Sec. 2 of Supplement 1). Interestingly, it has been pointed out that \cite{Osika2017, Yue2020, Parks2020}, including the variation of interband transition dipole moment in the saddle point analysis, can reveal a picture where electron and hole can ionize and recombine from different lattice site. It suggests that the present theory may have omitted some trajectories with non-zero spatial separation. Including these trajectories may lead to more critical points besides those in the dashed lines in Fig. \ref{figure02}.

	\begin{figure}[ht!]
		\centering
		\includegraphics[width=0.9\columnwidth]{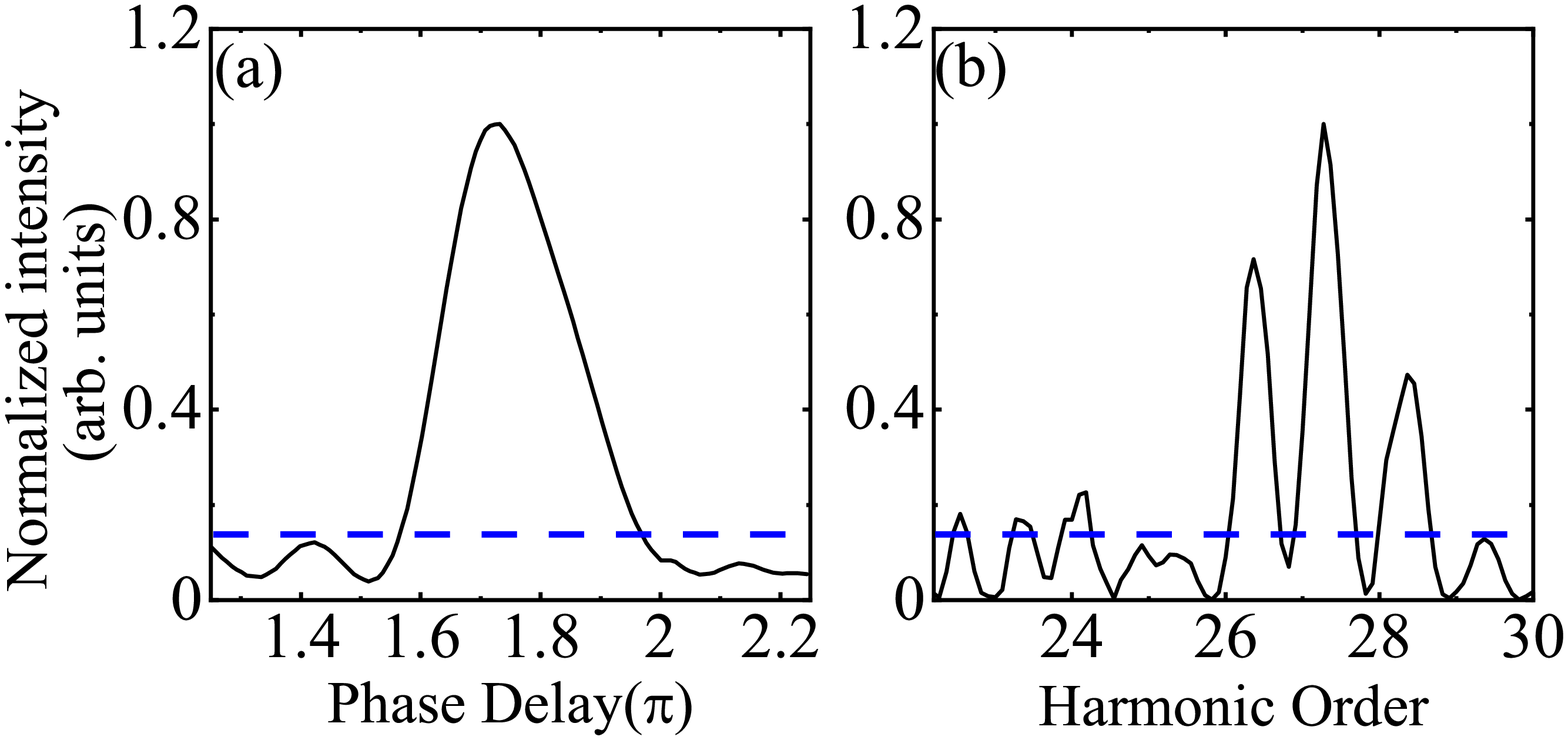}
		\caption{Sectional drawing near the swallowtail point in Fig. \ref{figure02}(b).
			(a) Intensity of the 27th order harmonic as a function of phase delay $\varphi$.
			(b) Intensity with respect to the harmonic orders at $\varphi=1.74\,\pi$.
			The maximum of the intensity is normalized to 1, while the intensity of the harmonics far away from the caustics zone is 0.14 (dashed lines) predicted by catastrophe theory.}
		\label{figure03}
	\end{figure}
	
	The relative enhancement of the caustic zone above has been given by catastrophe theory \cite{KravtsovSPU1983,RazNP2011}:
	\begin{equation}\label{Ienhan}
		I_{enhanced}/I_0\approx N^{2\delta},
	\end{equation}
	where $I_{enhanced}$ and $I_0$ are the intensity of the harmonic signal at and far away from the caustic zone, respectively. $N$ is a dimensionless constant used to describe the rapid oscillation phase in the steady phase approximation. In the case of HHG, $N$ is the harmonic order at the caustic zone. The value of $\delta$ is given according to the type of catastrophe. For the swallowtail type, it equals $3/10$, and thus, $I_{enhanced}/I_0\approx 7$. In Fig. \ref{figure03}, we show the enhancement effect of the harmonic signal in the caustic zone.	We plot the vertical section at the 27th order harmonic [see Fig. \ref{figure03}(a)] and the parallel section at $\varphi=1.74\, \pi$ [see Fig. \ref{figure03}(b)] for the swallowtail catastrophe in Fig. \ref{figure02}(b). According to \eqref{Ienhan}, the normalized intensity of the harmonic spectrum far away from the caustic zone is predicted to be 0.14, denoted by the dashed lines in Figs. \ref{figure03}(a) and \ref{figure03}(b), which are consistent with our simulations.

	The present calculations neglect the dephasing effect \cite{VampaPRL2014,KilenPRL2020}
	in the HHG process. In Fig. S1 we have also verified our analyses by taking account of dephasing time (see details in Sec. 2 of Supplement 1). We find that when the dephasing time is comparable to the laser optical period, e.g., half of the optical period, the above singular enhancement is weakened, but the spectral caustic structure remains. Specifically, since the green and blue trajectories in Fig. \ref{figure01}(c) are short, the caustic structures generating by the coalescence of the two branches remain sharp.

	In summary, we extended the HHG spectral caustic structure in atoms and molecules to solids in one dimension. We showed that besides the van Hove singularity determined by the local band structure, other singularities exist in the harmonic spectrum, which can be found by adjusting the parameters of the control space. Also, we obtained the diffraction pattern of the harmonic spectrum by solving the multielectron TDSE under the independent electron approximation. The structure and position of the pattern agree well with catastrophe theory and the semiclassical prediction. Our work is ready to extend to the multidimensional scenario, especially when the polarization direction of the laser field is along the symmetry axis of the solid lattice.  Further study is needed to explore the relationship between the caustic structures and other solid-state properties, such as multi-band dynamics, Berry curvatures, and transition dipole phases. The simple recollision model here needs to be generalized \cite{Osika2017, Yue2020, Parks2020} to treat more complicated and interesting semiclassical trajectories leading to different caustic patterns. Meanwhile, these caustic patterns may provide experimental evidence for the new models. Therefore, the present study is expected to open up an avenue to manipulate the harmonic spectrum structure and further enhance the harmonic yield by combining other techniques.

	\medskip
	
	\noindent\textbf{Funding.} This work was supported by the National Natural Science Foundation of China (Grants No. 11725417, No. 11974057), NSAF (Grant No. U1930403), and Science Challenge Project (Grant No. 2018005).
	
	\medskip
	
	\noindent\textbf{Disclosures.} The authors declare no conflicts of interest.
	
	\medskip
	
	\noindent\textbf{Data availability.} Data underlying the results presented in this paper are not publicly available at this time but may be obtained from the authors upon reasonable request.
	
	\medskip
	\noindent See Supplement 1 for supporting content.
	% Bibliography
	\bibliography{MyCollection}
	
	% Full bibliography added automatically for Optics Letters submissions; the following line will simply be ignored if submitting to other journals.
	% Note that this extra page will not count against page length
	\bibliographyfullrefs{MyCollection}
	
	%Manual citation list
	%\begin{thebibliography}{1}
	%\bibitem{Zhang:14}
	%Y.~Zhang, S.~Qiao, L.~Sun, Q.~W. Shi, W.~Huang, %L.~Li, and Z.~Yang,
	% \enquote{Photoinduced active terahertz metamaterials with nanostructured
	%vanadium dioxide film deposited by sol-gel method,} Opt. Express \textbf{22},
	%11070--11078 (2014).
	%\end{thebibliography}
	
	% Please include bios and photos of all authors for aop articles
	\ifthenelse{\equal{\journalref}{aop}}{%
		\section*{Author Biographies}
		\begingroup
		\setlength\intextsep{0pt}
		\begin{minipage}[t][6.3cm][t]{1.0\textwidth} % Adjust height [6.3cm] as required for separation of bio photos.
			\begin{wrapfigure}{L}{0.25\textwidth}
				\includegraphics[width=0.25\textwidth]{john_smith.eps}
			\end{wrapfigure}
			\noindent
			{\bfseries John Smith} received his BSc (Mathematics) in 2000 from The University of Maryland. His research interests include lasers and optics.
		\end{minipage}
		\begin{minipage}{1.0\textwidth}
			\begin{wrapfigure}{L}{0.25\textwidth}
				\includegraphics[width=0.25\textwidth]{alice_smith.eps}
			\end{wrapfigure}
			\noindent
			{\bfseries Alice Smith} also received her BSc (Mathematics) in 2000 from The University of Maryland. Her research interests also include lasers and optics.
		\end{minipage}
		\endgroup
	}{}

\end{document}